\documentclass[a4paper, 10pt, twocolumn]{article}

\usepackage[utf8]{inputenc}         
\usepackage[english]{babel}          

\usepackage{graphicx}               
\usepackage{tabularx}               
\usepackage{multirow}               
\usepackage{url}                

\usepackage[small,bf]{caption2}     
\usepackage{parskip}
\usepackage{titlesec}
\usepackage{amsmath}                
\usepackage{booktabs}
\usepackage{siunitx}
\usepackage[table]{xcolor} 
\usepackage{pgfplots} 
\usepackage{tikz}
\usepackage{collcell} 
\usepackage{dblfloatfix} 

\pgfplotsset{compat=1.18}

\pgfplotsset{
    colormap={corrmap}{
        rgb255(0cm)=(255,143,136) 
        rgb255(1cm)=(152,251,152) 
    }
}

\newcommand{\corrmin}{0.11}
\newcommand{\corrmax}{0.78}
\newcommand{\corralpha}{0.9} 

\newcommand{\applycorrcolor}[1]{%
\expandafter\applycorrcoloraux#1\relax
}
\def\applycorrcoloraux#1,#2,#3\relax{%
\pgfmathsetmacro{\rblend}{1-(1-#1)*\corralpha}
\pgfmathsetmacro{\gblend}{1-(1-#2)*\corralpha}
\pgfmathsetmacro{\bblend}{1-(1-#3)*\corralpha}
\edef\temp{\noexpand\cellcolor[rgb]{\rblend,\gblend,\bblend}}%
\temp%
}

\newcommand{\corr}[1]{%
\pgfmathsetmacro{\val}{#1}
\pgfmathsetmacro{\colorpos}{(\val-\corrmin)/(\corrmax-\corrmin)}
\pgfmathsetmacro{\colorpos}{max(0,min(1,\colorpos))}
\pgfplotscolormapaccess[0:1]{\colorpos}{corrmap}
\applycorrcolor{\pgfmathresult}#1
}

\titleformat{\section}{\normalfont\large\bfseries}{\thesection}{}{}
\titleformat{\subsection}{\normalfont\large\bfseries}{\thesection}{}{}
\titleformat{\paragraph}{\normalfont\bfseries}{\theparagraph}{}{}
\titlespacing{\section}{0pt}{6pt}{-1pt}
\titlespacing{\subsection}{0pt}{3pt}{-1pt}
\titlespacing{\paragraph}{0pt}{3pt}{-1pt}

\newcolumntype{Y}{>{\centering\arraybackslash}X}    

\let\OLDthebibliography\thebibliography
\renewcommand\thebibliography[1]{%
  \OLDthebibliography{#1}%
  \setlength{\itemsep}{0pt}%
  \setlength{\parskip}{0pt}%
}

\begin{document}

\date{}                                         

\title{\vspace{-8mm}\textbf{\large
Embedding-Based Intrusive Evaluation Metrics for Musical Source Separation Using MERT Representations}}

\author{
Paul A. Bereuter$^1$, Alois Sontacchi$^1$\\
$^1$ \emph{\small University of Music and Performing Arts Graz, Institute of Electronic Music and Acoustics (IEM)}\\
\emph{\small A-8010 Graz, Austria, Email: \{bereuter,sontacchi\}@iem.at}
}\maketitle
\thispagestyle{empty}           
\section*{Abstract}
Evaluation of musical source separation (MSS) has traditionally relied on Blind Source Separation Evaluation (BSS-Eval) metrics. However, recent work suggests that BSS-Eval metrics exhibit low correlation between metrics and perceptual audio quality ratings from a listening test, which is considered the gold standard evaluation method. As an alternative approach in singing voice separation, embedding-based intrusive metrics that leverage latent representations from large self-supervised audio models such as Music undERstanding with large-scale self-supervised Training (MERT) embeddings have been introduced. In this work, we analyze the correlation of perceptual audio quality ratings with two intrusive embedding-based metrics: a mean squared error (MSE) and an intrusive variant of the Fréchet Audio Distance (FAD) calculated on MERT embeddings. Experiments on two independent datasets show that these metrics correlate more strongly with perceptual audio quality ratings than traditional BSS-Eval metrics across all analyzed stem and model types.
\section*{Introduction}
The gold standard in evaluating musical source separation models is the perceptual listening test. Test formats such as the Multiple Stimuli with Hidden Reference and Anchor (MUSHRA) test \cite{ITU-R-BS.1534-3}, the Absolute Category Rating (ACR) or the Degradation Category Rating (DCR) test \cite{ITU-P808-2021} are commonly used to obtain perceptual quality ratings. However, in order to obtain meaningful perceptual ratings one requires a substantial number of audio samples and a meaningful sample size of ratings per  sample. For multiple MSS models and configurations this can easily get costly and time-consuming. Therefore, objective evaluation metrics, whose computation requires only a fraction of the time and effort of listening tests, yet correlate well with perceived audio quality, can act as perceptual proxies for such tests. The most commonly used objective evaluation metrics in musical source separation are the Blind Source Separation Evaluation (BSS-Eval) metrics \cite{vincent2006}. In a traditional musical source separation scenario, the musical mixtures are separated into four sources (stems): vocals, bass, drums and other instruments. To get a global quality measure the BSS-Eval metrics project the estimation error onto the target signal subspace. The projected overall error (distortion) yields the global quality measure interpretable as an energy ratio between the target signal and the distortion, called Signal-to-Distortion Ratio (SDR). The overall error can further be decomposed into an error term quantifying the interferences from other sources and artifacts (e.g. unwanted non-linearities or signal dynamics), resulting in the Signal-to-Interference Ratio (SIR) and the Signal-to-Artifact Ratio (SAR). In \cite{sisdrRoux2019} scale-invariant versions of these metrics (SI-SDR, SI-SAR \& SI-SIR) were proposed. 
More recently several works have shown, that these BSS-Eval metrics do not correlate well with perceptual audio quality ratings \cite{cano2016, ward2018, torcoli2021}. In our previous work \cite{bereuter2025gensvs} we showed that among other evaluation metrics, BSS-Eval metrics exhibit poor correlation with perceptual audio quality ratings, especially for novel generative singing voice separation models.
We provided an alternative with the usage of embedding-based intrusive metrics, which leverage latent representations form large-scale self-supervised audio encoder models, which showed increased correlation with perceptual quality labels across both discriminative and generative model types. 

In this work we will investigate two embedding-based intrusive metrics, calculated on MERT embeddings \cite{li2024mert}. The Music understanding with large-scale self-supervised Training (MERT) model is a transformer encoder architecture that uses a self-supervised learning paradigm with teacher models, that allow for capturing both acoustics and musical properties in low dimensional contextual embeddings. In \cite{bereuter2025gensvs} we analyzed the correlation of said MERT-based evaluation metrics with audio quality ratings for singing voice separation models. This work extends this previous analysis with the correlation for four stem types of standard MSS models (vocals, bass, drums and other) on an independent dataset. The next sections provide details on the calculation of the two embedding-based metrics under investigation, the baseline metrics we compare against, and outline the existing datasets we use for the correlation analysis. Finally, we present the results of the correlation analysis and discuss the implications of our findings. 
\section*{Embedding-Based Evaluation}
The metrics under investigation are computed on MERT embeddings \cite{li2024mert} of the target and the separated audio signal, obtained by passing the respective audio signals through a pre-trained MERT encoder and extracting the latent representations at a specific layer. We denote the MERT encoder as a function $f_{\mathrm{MERT}}^{l}()$ with $l$ being the layer index. We use the MERT-v95 model \cite{li2024mert} and extract the embeddings at layer 12 and denote the embeddings of the target signal $\boldsymbol{x}$ and the separated audio signal $\hat{\boldsymbol{x}}$ as $\boldsymbol{E} = f_{\mathrm{MERT}}^{12}(\boldsymbol{x})$ and $\hat{\boldsymbol{E}} = f_{\mathrm{MERT}}^{12}(\hat{\boldsymbol{x}})$.
The first metric we investigate is the mean squared error denoted in equation (\ref{eq:mert_mse}), calculated between the MERT embeddings of the target and separated audio signal ($\mathrm{MSE}_{\mathrm{MERT}}$).
\begin{equation}
\mathrm{MSE}_{\mathrm{MERT}}= \frac{1}{NM}\left\lVert \boldsymbol{E} - \hat{\boldsymbol{E}} \right\rVert_F^2
\label{eq:mert_mse}
\end{equation}
The MERT model operates on audio with a sampling frequency of \SI{24000}{\hertz}, which for \SI{5}{\second} audio chunks results in a representation with the dimensionality of $N=768$ embeddings and $M=\mathrm{374}$ time frames.

The second embedding-based metric we investigate is an intrusive variant of the Fréchet Audio Distance (FAD). The FAD \cite{kilgour19_interspeech} was proposed to assess the quality of generative audio using non-matching references. This means that the FAD as described in equation (\ref{eq:mert_fad}) is calculated between audio embeddings of high-quality reference samples (reference distribution), which do not have to be exact targets for the embeddings of generated audio samples (test distribution). The distance measure is computed between the multivariate mean and covariance of the target ($\boldsymbol{\mu}$ and $\boldsymbol{\Sigma}$) and test distribution ($\hat{\boldsymbol{\mu}}$ and $\hat{\boldsymbol{\Sigma}}$). By characterizing the embedding distributions solely through their empirical mean and covariance, the metric implicitly approximates the embedding distributions as multivariate Gaussian distributions. 

\begin{equation}
\resizebox{\dimexpr\columnwidth-2.5em\relax}{!}{$
\mathrm{FAD}
= \left\lVert \boldsymbol{\mu} - \hat{\boldsymbol{\mu}} \right\rVert_2^2
+ \operatorname{tr}\!\left(
\boldsymbol{\Sigma}
+ \hat{\boldsymbol{\Sigma}}
- 2\left(\boldsymbol{\Sigma}\hat{\boldsymbol{\Sigma}}\right)^{1/2}
\right)
$}
\label{eq:mert_fad}
\end{equation}

In our case, we use an intrusive variant of FAD in which the mean and covariance are computed individually for the target and separated signals of each song. The temporal sequences of embeddings of the target and separated signals are treated as samples drawn from the reference and test distributions, respectively. In \cite{gui2024adapting}, this per-song FAD variant ($\mathrm{FAD}_{\mathrm{song2song}}$) was used for outlier detection within datasets. In \cite{bereuter2025gensvs}, we showed that it also works well for the quality assessment of singing voice separation models, as it exhibited good correlation with perceptual audio quality ratings. 
We calculate the embedding-based evaluation metrics using the \texttt{gensvs} Python package\footnote{https://pypi.org/project/gensvs/} we presented in \cite{bereuter2025gensvs}.
\section*{Baseline Metrics}
We compare the embedding-based metrics against the overall quality metric SDR and the scale-invariant BSS-Eval metrics SI-SDR, SI-SAR and SI-SIR. 
To compute SDR and SI-SDR we use the \texttt{torchmetrics} Python package\footnote{https://pypi.org/project/torchmetrics/} and for SI-SAR and SI-SIR we use the implementation of the \texttt{nussl} library\footnote{https://github.com/nussl/nussl}. Additionally, we include a spectral mean squared error (MSE$_\text{spec}$) calculated between the magnitude values of the target and separated signal's STFT. To calculate the STFT we used a Hann window with a window length of $N_{\mathrm{FFT}}=512$ samples and a hop-size of $R=256$ samples.
\section*{Perceptual Audio Quality Ratings}
We use two existing datasets that contain perceptual audio quality ratings of musical and singing voice separation models alongside audio samples on which the presented metrics can be computed. We reexamine our previous results from \cite{bereuter2025gensvs} (GenSVS dataset) and extend our evaluation with a correlation analysis conducted on the dataset published in \cite{jaffe2025bakeoff} (Bake-Off dataset).
\subsection*{Bake-Off Dataset\footnote{https://zenodo.org/records/15843081}}
The dataset presented in \cite{jaffe2025bakeoff} contains perceptual audio quality ratings for four-stem musical source separation models applied to \SI{30}{\second} MUSDB18-HQ test-set from the 2018 Signal Separation Evaluation Campaign \cite{sigsep2018}. We use the presented results for the 3 NN-based discriminative MSS models and the ideal ratio mask (IRM1), which was included as an oracle method in the listening test presented in \cite{jaffe2025bakeoff}. The ratings were collected for the first \SI{10}{\second} of the test-set using a MUSHRA test \cite{ITU-R-BS.1534-3}. We used the same quality check conditions as proposed in \cite{jaffe2025bakeoff} and allowed for two quality check violations to screen the dataset for outliers. The individual ratings were averaged to 800 MUSHRA scores (4 models $\times$ 4 stem types $\times$ 50 songs), ranging from 0 (bad) to 100 (excellent). The \SI{10}{\second} target and IRM1 audio samples are not included in the dataset accompanying \cite{jaffe2025bakeoff}. However, they can be extracted from the dataset\footnote{https://zenodo.org/records/1256003} which accompanies the original 2018 Signal Separation Evaluation Campaign \cite{sigsep2018}.
\subsection*{GenSVS Dataset \footnote{https://zenodo.org/records/15911723}}
In our previous work \cite{bereuter2025gensvs} we collected audio quality ratings for 3 discriminative and 2 generative singing voice separation models, which were applied to \SI{5}{\second} excerpts of the MUSDB18-HQ \cite{musdb18} test-set. The ratings were collected using a ITU P.808 standardized DCR test \cite{ITU-P808-2021}, which is a listening test paradigm, that asks for the evaluation of degradation in audio quality compared to a reference. We were able to aggregate 12 ratings per audio samples which were averaged into a total of 250 degradation mean opinion score (DMOS), ranging from 1 (\grqq{}degradation is very annoying\grqq{}) to 5 (\grqq{}degradation is inaudible\grqq{}). More information on the underlying dataset incl. data screening process can be found in \cite{bereuter2025gensvs}.

\subsection*{Correlation Analysis}
To investigate whether any of the presented metrics can act as a reasonable proxy for the perceptual audio quality ratings, we calculate Spearman's rank correlation coefficient (SRCC) to measure the monotonicity in the relation of objective evaluation metrics and perceptual audio quality ratings \cite{spearman1904}. To quantify the linearity in the relation between the metrics and the perceptual ratings we also calculate Pearson's correlation coefficient (PCC) \cite{pearson1896}. We calculate these correlation coefficients for the pooled data within each stem type and for the overall pooled data across all stem types. For the GenSVS dataset we calculate the correlation coefficients separately for the generative and discriminative model types, next to the overall pooled correlation.
\section*{Results}
\begin{figure}[htbp]
    \begin{center}
        \includegraphics[width=\columnwidth]{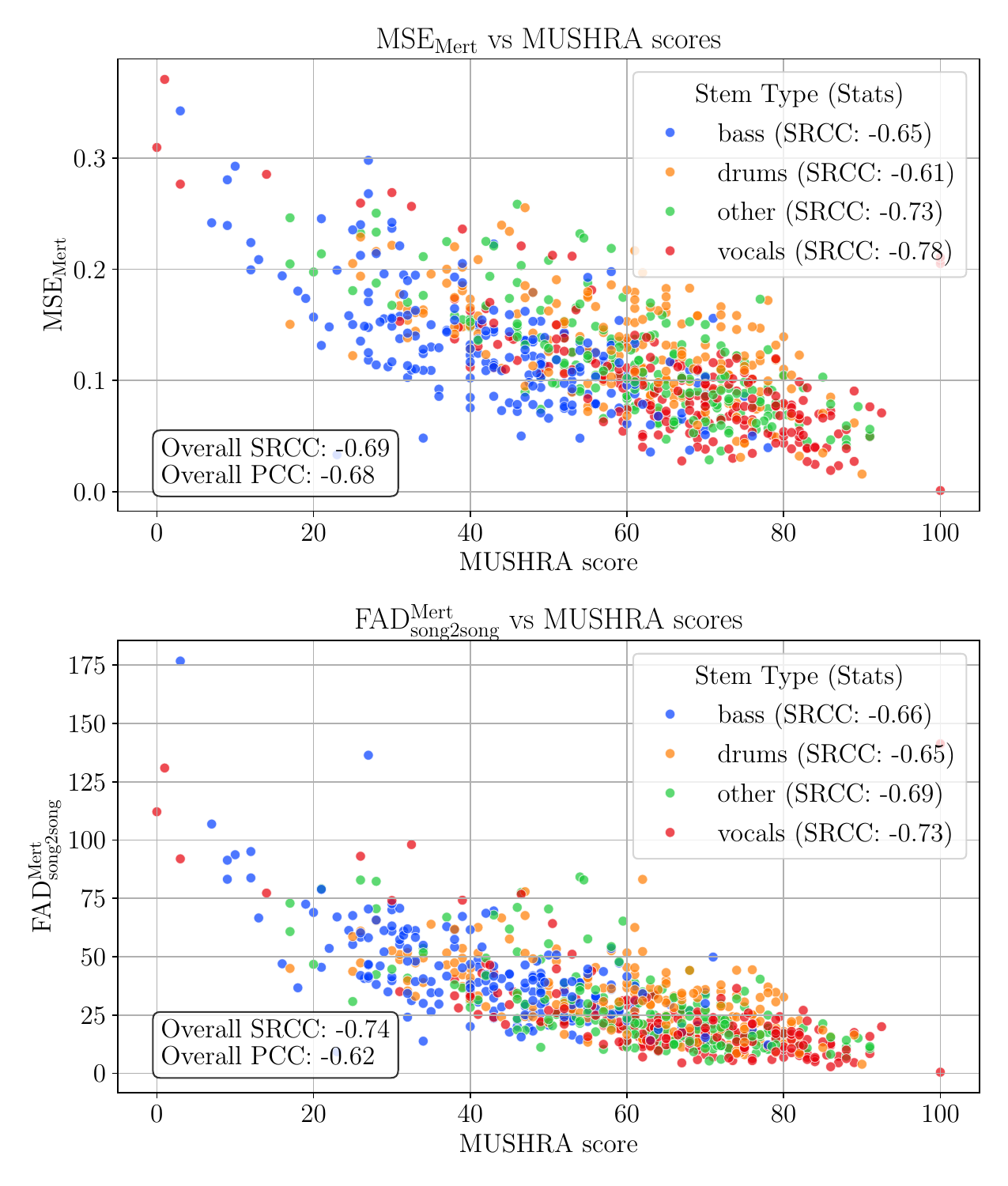}
    \end{center}
    \caption{Scatter plots between the both embedding-based metrics and the MUSHRA scores for the Bake-Off dataset. Different stem types are indicated by different colors.}
    \label{fig:scatter}
    \vspace{-0.6cm}
\end{figure}

The correlation coefficients for the Bake-Off dataset are presented in Table \ref{tab:correlations_bakeoff}, while the results for the GenSVS dataset are shown in Table \ref{tab:correlations_gensvs}. Higher correlation values are highlighted in green, whereas lower values are marked in red.
The highest correlations are observed for the embedding-based metrics for the vocal stem type. Thus, the results reported for the GenSVS dataset from \cite{bereuter2025gensvs}, summarized in Table \ref{tab:correlations_gensvs} can be reproduced on a second independent dataset. Among the BSS-Eval metrics, the highest correlations on the Bake-Off dataset are obtained for SDR and SI-SAR on the vocal stems. This observation is consistent with the findings reported in \cite{jaffe2025bakeoff}. A similar trend can also be observed for the GenSVS dataset, although only for discriminative models. As the Bake-Off dataset contains only audio samples generated by discriminative models, similar results for SI-SAR and SDR are observed across both datasets. Only for $FAD_{song2song}^{MERT}$, PCC and SRCC are not aligned for the vocal stem type in the Bake-Off dataset. Here the PCC being lower than the SRCC suggests a monotonic but not necessarily linear relationship between this metric and the perceptual audio quality ratings. This also influences the overall correlation of the Bake-Off dataset in Table \ref{tab:correlations_bakeoff} where the overall PCC is also lower than the SRCC. This becomes visible when looking at the scatter plot between $FAD_{song2song}^{MERT}$ and the MUSHRA scores depicted in Figure \ref{fig:scatter}. Here a slightly curved relationship between the metric and the ratings towards the lower end of the rating scale is noticeable. Figure \ref{fig:scatter} also shows that the separated bass stems achieved the lowest ratings in the listening test, as most data points lie below a MUSHRA score of 60. The lower correlation values for bass stems in both embedding-based and BSS-Eval metrics suggest that the metrics struggle to capture the variance toward the lower end of the rating scale. Nonetheless, the embedding-based metrics still achieve correlation coefficients above 0.65 for bass stems, indicating a stronger correlation than all baseline metrics. When looking at the rank-based correlation we see a higher overall SRCC for the FAD metric, suggesting a more monotonic relationship when pooling all stem types.

When looking at the results in Table \ref{tab:correlations_bakeoff} and \ref{tab:correlations_gensvs} on a higher level the most notable result, is that across all stem types and for both datasets, the embedding-based metrics consistently outperform the BSS-Eval metrics. Also, MSE$_\text{spec}$ shows inferior correlation results on both datasets, which is expectable, as this metric does not account for any perceptual aspects.

\section*{Conclusion}
This study investigates the correlation between two novel embedding-based and established baseline metrics with perceptual audio quality labels on the Bake-Off \cite{jaffe2025bakeoff} and GenSVS dataset \cite{bereuter2025gensvs}. The results show that the proposed embedding-based metrics consistently achieve the highest correlations with human ratings across all stem types for the evaluated systems. In particular, the strong correlations observed for vocal stems confirm the findings previously reported for the GenSVS dataset \cite{bereuter2025gensvs} and demonstrate that these results generalize to the independently created Bake-Off dataset.
Among traditional BSS-Eval metrics SDR and SI-SAR exhibit the highest correlations with perceptual ratings for vocal stems, however only for discriminative singing voice separation models, as the correlation analysis on the GenSVS dataset shows. The lowest correlations between metrics and perceptual audio quality ratings are observed for the spectral MSE, which is a purely signal-based metric.

Overall, the results suggest that embedding-based metrics provide a more reliable perceptual proxy for evaluating musical source separation models than traditional BSS-Eval metrics. This is observed for both discriminative and generative singing voice separation models as well as across all stem types in a classical four-stem separation scenario. Future work could investigate if such embedding-based evaluation methods also generalize well to a broader range of source separation models such as generative source separation models. In addition, the rapidly growing number of audio foundation models provides a wide range of neural audio representations that may more accurately capture the nuances of perceptual audio quality nuances relevant to musical source separation. Finally, the results suggest that selecting adequate metrics for different evaluation scenarios that evaluate separation performance for specific subclasses, such as model types or stem types, can provide a perceptually well-founded and more discriminative evaluation than aggregated \grqq{}one-metric-fits-all\grqq{} evaluations. All code to calcualte the metrics and execute the correlation analysis is available at \url{https://github.com/pablebe/mert-emb-eval/}
\begin{table*}[t!]
\centering
\small
\setlength{\tabcolsep}{3pt}
\caption{SRCC and PCC between metrics and ratings for the Bake-Off dataset, evaluated for stem-type and overall data pooling. The correlation coefficients for the MSE and FAD metrics were multiplied with -1 for comparability. Green indicates high correlation, red indicates low correlation}
\label{tab:correlations_bakeoff}

\resizebox{\textwidth}{!}{%
\begin{tabular}{l*{14}{c}}
\toprule
& \multicolumn{2}{c}{MSE$_{\text{Mert}}$}
& \multicolumn{2}{c}{FAD$^{\text{Mert}}_{\text{song2song}}$}
& \multicolumn{2}{c}{SDR}
& \multicolumn{2}{c}{SI-SDR}
& \multicolumn{2}{c}{SI-SAR}
& \multicolumn{2}{c}{SI-SIR}
& \multicolumn{2}{c}{MSE$_{\text{spec}}$} \\
\cmidrule(lr){2-3} \cmidrule(lr){4-5} \cmidrule(lr){6-7} \cmidrule(lr){8-9}
\cmidrule(lr){10-11} \cmidrule(lr){12-13} \cmidrule(lr){14-15}
Stem & SRCC & PCC & SRCC & PCC & SRCC & PCC & SRCC & PCC & SRCC & PCC & SRCC & PCC & SRCC & PCC \\
\midrule

Bass
& \corr{0.65} & \corr{0.67}
& \corr{0.66} & \corr{0.67}
& \corr{0.43} & \corr{0.48}
& \corr{0.47} & \corr{0.52}
& \corr{0.51} & \corr{0.55}
& \corr{0.27} & \corr{0.30}
& \corr{0.39} & \corr{0.38} \\

Drums
& \corr{0.61} & \corr{0.62}
& \corr{0.65} & \corr{0.65}
& \corr{0.31} & \corr{0.31}
& \corr{0.35} & \corr{0.35}
& \corr{0.39} & \corr{0.39}
& \corr{0.22} & \corr{0.22}
& \corr{0.28} & \corr{0.11} \\

Other
& \corr{0.73} & \corr{0.76}
& \corr{0.69} & \corr{0.68}
& \corr{0.62} & \corr{0.67}
& \corr{0.65} & \corr{0.69}
& \corr{0.61} & \corr{0.64}
& \corr{0.37} & \corr{0.37}
& \corr{0.32} & \corr{0.22} \\

Vocals
& \corr{0.78} & \corr{0.78}
& \corr{0.73} & \corr{0.52}
& \corr{0.69} & \corr{0.71}
& \corr{0.69} & \corr{0.68}
& \corr{0.70} & \corr{0.68}
& \corr{0.59} & \corr{0.64}
& \corr{0.54} & \corr{0.51}\\

Overall
& \corr{0.69} & \corr{0.68}
& \corr{0.74} & \corr{0.62}
& \corr{0.46} & \corr{0.48}
& \corr{0.49} & \corr{0.44}
& \corr{0.49} & \corr{0.44}
& \corr{0.35} & \corr{0.37}
& \corr{0.37} & \corr{0.30} \\

\bottomrule
\end{tabular}
}
\end{table*}

\begin{table*}[t!]
\centering
\small
\setlength{\tabcolsep}{3pt}
\caption{SRCC and PCC between metrics and ratings for the GenSVS dataset, evaluated for model-type (discriminative and generative models) and overall data pooling. The correlation coefficients for the MSE and FAD metrics were multiplied with -1 for comparability. Green indicates high correlation, red indicates low correlation.}
\label{tab:correlations_gensvs}
\resizebox{\textwidth}{!}{%
\begin{tabular}{lcccccccccccccc}
\toprule
& \multicolumn{2}{c}{MSE$_{\text{Mert}}$} & \multicolumn{2}{c}{FAD$^{\text{Mert}}_{\text{song2song}}$} & \multicolumn{2}{c}{SDR} & \multicolumn{2}{c}{SI-SDR} & \multicolumn{2}{c}{SI-SAR} & \multicolumn{2}{c}{SI-SIR} & \multicolumn{2}{c}{MSE$_{\text{spec}}$} \\
\cmidrule(lr){2-3} \cmidrule(lr){4-5} \cmidrule(lr){6-7} \cmidrule(lr){8-9} \cmidrule(lr){10-11} \cmidrule(lr){12-13} \cmidrule(lr){14-15}
Stem & SRCC & PCC & SRCC & PCC & SRCC & PCC & SRCC & PCC & SRCC & PCC & SRCC & PCC & SRCC & PCC \\
\midrule
Vocals (gen. models) & \corr{0.71} & \corr{0.77} & \corr{0.65} & \corr{0.69} & \corr{0.18} & \corr{0.22} & \corr{0.11} & \corr{0.09} & \corr{0.11} & \corr{0.08} & \corr{0.25} & \corr{0.35} & \corr{0.41} & \corr{0.49} \\
Vocals (disc. models) & \corr{0.76} & \corr{0.75} & \corr{0.62} & \corr{0.62} & \corr{0.68} & \corr{0.62} & \corr{0.69} & \corr{0.54} & \corr{0.69} & \corr{0.53} & \corr{0.57} & \corr{0.59} & \corr{0.53} & \corr{0.46} \\
Overall & \corr{0.67} & \corr{0.70} & \corr{0.60} & \corr{0.61} & \corr{0.24} & \corr{0.18} & \corr{0.25} & \corr{0.09} & \corr{0.24} & \corr{0.08} & \corr{0.36} & \corr{0.39} & \corr{0.41} & \corr{0.41} \\
\bottomrule
\end{tabular}
}
\end{table*}
\bibliographystyle{IEEEtran}
\bibliography{refs25}
\end{document}